\documentclass{article}

\usepackage{arxiv}

\usepackage[utf8]{inputenc} 
\usepackage[T1]{fontenc}    
\usepackage{hyperref}       
\usepackage{url}            
\usepackage{booktabs}       
\usepackage{amsfonts}       
\usepackage{nicefrac}       
\usepackage{microtype}      
\usepackage{lipsum}
\usepackage{graphicx}
\graphicspath{ {./images/} }
\usepackage{natbib}
\usepackage{amsmath}
\usepackage{multirow}

\title{ERA-IT: Aligning Semantic Models with Revealed Economic Preference for Real-Time and Explainable Patent Valuation}

\author{
 Yongmin Yoo \\
  School of Computing\\
  Macquarie University\\
  Sydney, NSW 2109\\
  Australia\\
  \texttt{yooyongmin91@gmail.com} \\
 \And
 Seungwoo Kim \\
  School of Computer Science\\
  University of Technology Sydney\\
  Sydney, NSW 2007\\
  Australia\\
  \texttt{fo.swkim@gmail.com} \\
 \And
 Jingjiang Liu \\
  School of Management\\
  Zhejiang University\\
  Hangzhou, Zhejiang 310058\\
  China\\
  \texttt{liujingjiang@zju.edu.cn} \\
}

\begin{document}
\maketitle
\begin{abstract}
This study proposes the Economic Reasoning Alignment via Instruction Tuning (ERA-IT) framework, which aligns the semantic reasoning of Large Language Models (LLMs) with revealed economic preferences to establish a new paradigm for real-time and interpretable patent valuation. We theoretically conceptualize patent renewal history as a revealed economic preference and leverage it as an objective supervisory signal to align the generative reasoning of LLMs with market realities, a process we term Eco-Semantic Alignment. Our empirical analysis of 10,000 randomly sampled European Patent Office patents across diverse technological domains confirms the efficacy of this approach. The ERA-IT framework achieves superior performance across all evaluated metrics compared to the baseline models. Specifically, it surpasses the traditional Random Forest model based on TF-IDF features by 36.3\% in accuracy, 33.1\% in Macro-F1, and 88.1\% in the Matthews Correlation Coefficient (MCC). Against the discriminative pre-trained language model Longformer, ERA-IT exhibits gains of 11.5\% in accuracy, 8.7\% in Macro-F1, and 27.4\% in MCC. It also exceeds the zero-shot performance of the generative LLM GPT-5-mini by 9.6\% in accuracy, 8.3\% in Macro-F1, and 21.5\% in MCC. The ablation study confirms the criticality of the Rationale Generation module (Economic Chain-of-Thought), whose removal led to the largest performance decline, with a 6.4\% drop in Macro-F1. Moreover, ERA-IT demonstrates robust applicability across heterogeneous patent landscapes.
\end{abstract}
\section{Introduction}

\begin{figure}[ht]
    \centering
    \includegraphics[width=0.8\textwidth]{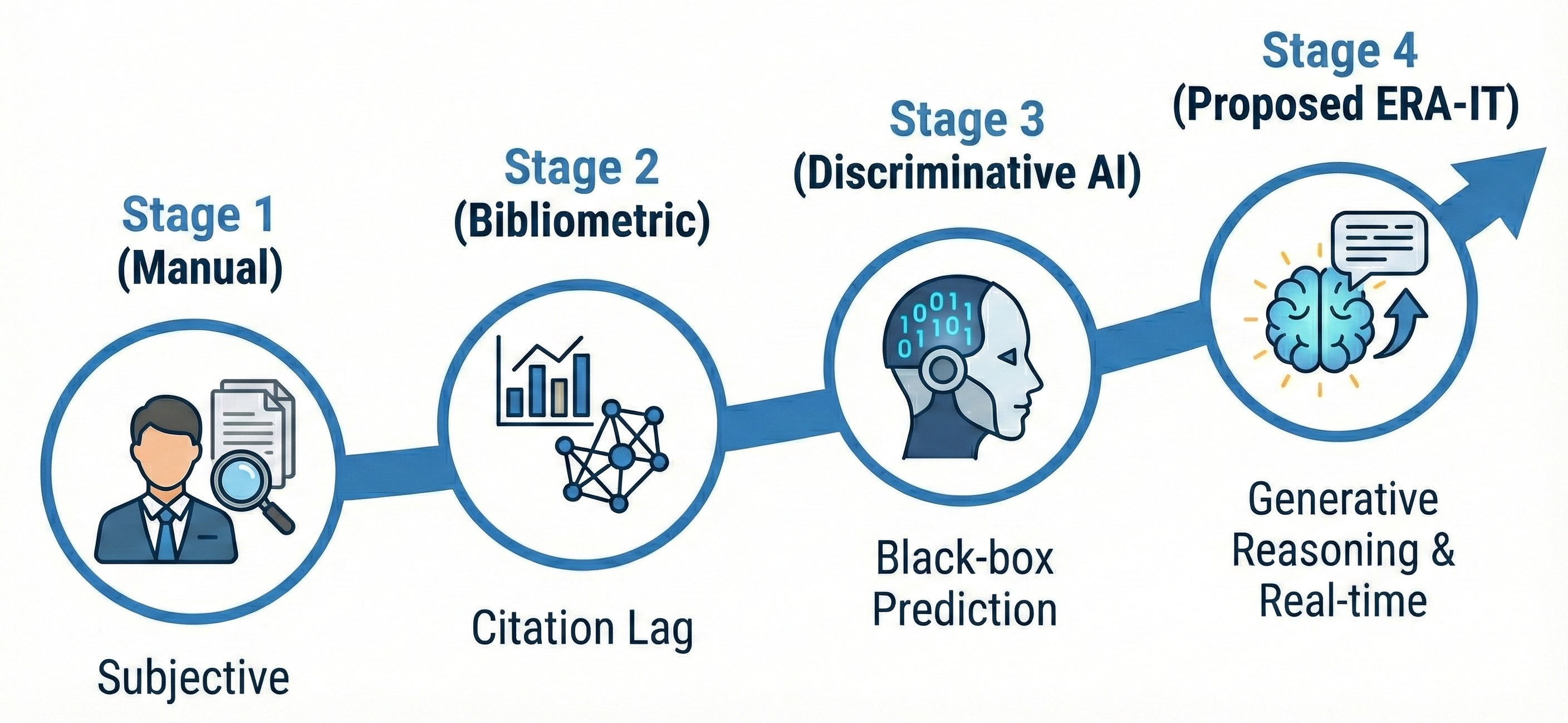}
    \caption{Evolutionary trajectory of patent valuation methodologies. The figure illustrates the paradigm shift from subjective manual assessment and latency-prone bibliometric indicators to the proposed ERA-IT framework, which enables real-time, explainable valuation via generative reasoning.}
    \label{fig:valuation_evolution}
\end{figure}

In the contemporary knowledge-driven economy, patents, as core intangible assets, are increasingly pivotal to firms' strategic decisions and value creation~\citep{ernst2003}. They serve as the fundamental currency of innovation, delineating a firm's technological trajectory and securing market dominance~\citep{hall2005, reitzig2004}. Accurate and timely patent valuation forms the bedrock of firms' strategy for managing technological innovation~\citep{fischer2014, kalip2022}. However, achieving objective and accurate ex-ante valuation of patents remains a classic and persistent grand challenge in information science, information economics, and technological innovation management~\citep{reitzig2004}. Figure~\ref{fig:valuation_evolution} outlines the evolutionary path of patent valuation methodologies.

The crux of this challenge stems from two critical dimensions. First, patents themselves are complex and information-dense hybrid ``technical-legal'' artifacts. Their true value is deeply encoded in unstructured technical texts~\citep{han2015}. This creates profound information asymmetry between informed patent holders and external evaluators, including investors and managers, all of whom struggle to penetrate this ``technical black box'' and transact efficiently in markets for technology~\citep{akerlof1970, gambardella2007, gans2008}. This opacity creates severe market friction, where stakeholders struggle to distinguish high-value inventions from the vast noise of global filings due to the high-dimensional complexity of mapping technical specifications to market outcomes~\citep{pakes1986, trajtenberg1990}. Second, information asymmetry is inherent in the global patent ecosystem. Unlike domain-specific studies that benefit from homogeneous terminologies, a generalized valuation framework must contend with heterogeneous linguistic patterns across disparate technological sectors~\citep{pakes1984}. In this context, traditional bibliometric indicators, such as citation counts, fail to serve as effective signals for timely decision-making due to systemic temporal latency; a patent typically accumulates meaningful citation data only years after its publication~\citep{harhoff2003}. This ``backward-looking'' nature of bibliometrics prevents managers from assessing patent value in a way that is both real-time and closely reflects genuine market preferences. Consequently, there exists an urgent imperative to bridge the gap between technical disclosure and economic realization. This imperative drives the quest for advanced architectures capable of leveraging ``forward-looking'' signals, constituting a critical research frontier. 

In this context, historical data on patent renewal has gained theoretical prominence as a direct signal of ``revealed economic preferences,'' since the act of renewal represents a ``vote with the wallet'' that is more closely aligned with market reality than any external proxy~\citep{pakes1986, pakes1989, schankerman1986}. From the perspective of economic signaling theory~\citep{jiang2026}, the decision to pay monotonically increasing renewal fees reveals the patent holder's preference, indicating that the asset's private value exceeds its maintenance expenses~\citep{danish2020,jou2018,porter2023,kumar2024}. Unlike stochastic citations, renewal records provide a definitive, discrete signal of value maintenance available throughout the patent's lifecycle. However, a fundamental methodological gap persists: how to effectively align this structured, low-dimensional economic-behavior signal with the high-dimensional, unstructured semantic information in patent texts to build an accurate and explainable predictive model. Existing frameworks have struggled to integrate this objective economic signal with the rich, unstructured semantic information embedded within patent documents~\citep{arts2018}. Traditional Natural Language Processing (NLP) techniques often fail to capture the nuanced causal relationships between the specific exclusionary scope defined in the claims and the patent's potential for economic longevity~\citep{lee2020}.

The recent advent of Large Language Models (LLMs) offers a transformative opportunity to overcome these linguistic barriers~\citep{yan2025}. LLMs possess the unique capacity for complex reasoning and for extracting latent semantic features. Nevertheless, applying general-purpose LLMs to patent valuation is non-trivial; these models often lack the domain-specific intuition required to align technical novelty with economic viability. To overcome this critical barrier in high-stakes Intellectual Property (IP) management, there is an urgent need for adaptive technological and methodological innovations. The development of a predictive architecture that effectively fuses the objective ground truth of renewal data with the semantic reasoning capabilities of LLMs, a process that we term ``Eco-Semantic Alignment,'' represents a largely unexplored frontier in intelligent information systems.

\begin{figure}[ht]
    \centering
    \includegraphics[width=0.8\textwidth]{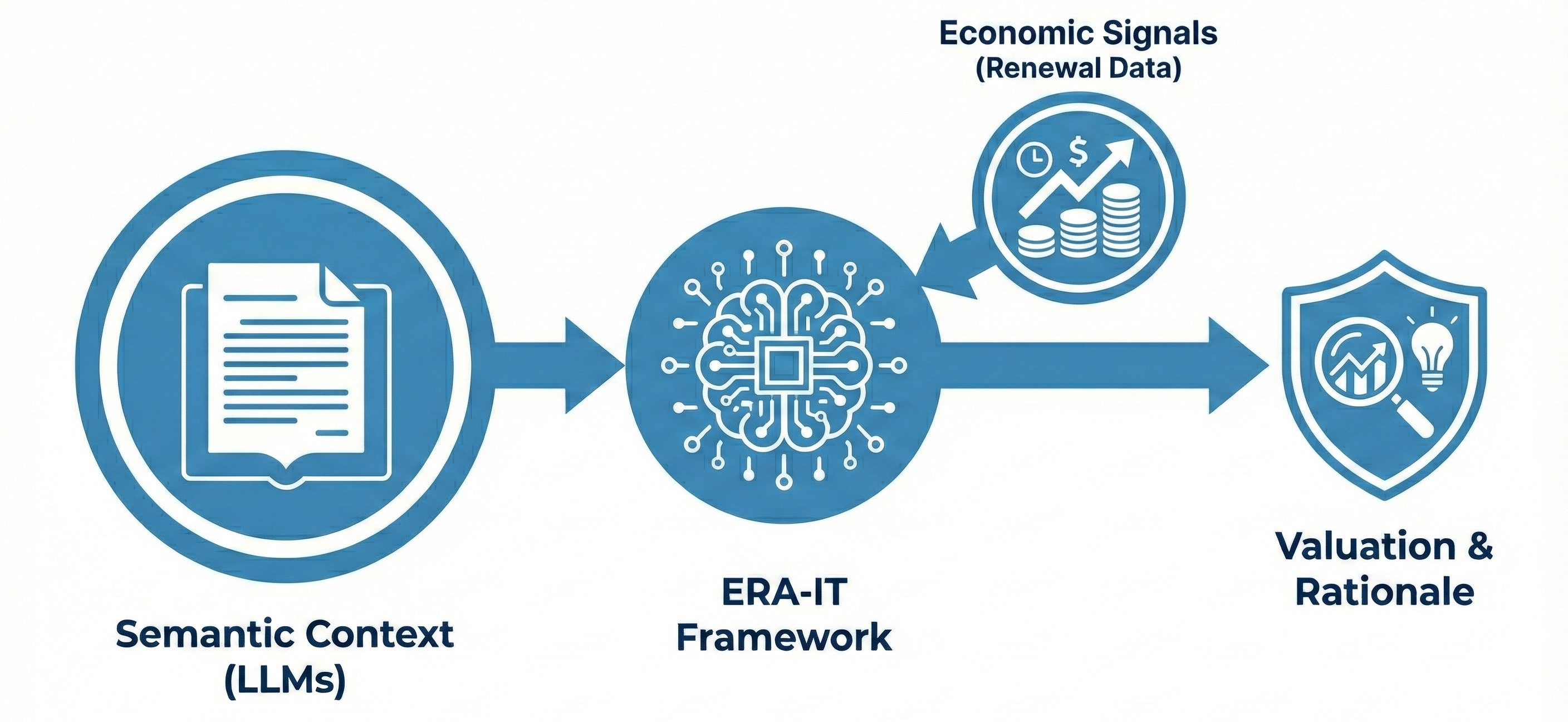}
    \caption{Conceptual overview of the ERA-IT framework. The architecture integrates high-dimensional semantic context processed by LLMs with objective economic signals derived from renewal data to generate explainable valuation outcomes.}
    \label{fig:concept_overview}
\end{figure}

To bridge this methodological gap, this study proposes the Economic-Reasoning Alignment via Instruction Tuning (ERA-IT) framework, as illustrated in Figure~\ref{fig:concept_overview}. Our approach uses European Patent Office (EPO) renewal data as a supervisory signal to guide LLMs' generative reasoning. Guided by a rigorous data fusion strategy, we operationalize the longitudinal payment of renewal fees as a value proxy. This allows us to translate economic behavior into a labeled dataset with discrete value tiers, transforming the ambiguous task of valuation into a robust supervised learning problem. Subsequently, we employ a strategy that synergizes parameter-efficient fine-tuning with chain-of-thought prompting. This alignment enables the model to learn latent semantic patterns correlated with economic survival, effectively converting unstructured text into quantifiable predictive indicators.

To rigorously validate the efficacy and generalizability of this framework, we conducted a comprehensive empirical analysis using a dataset comprising over 10,000 EPO patents, randomly sampled across various International Patent Classification (IPC) sections. This randomized sampling strategy ensures that our findings are not biased toward specific high-velocity domains and instead represent a robust solution applicable to the broader technological landscape. Our results demonstrate that this hybrid approach yields significantly superior predictive accuracy compared to conventional rule-based heuristics and statistical baselines~\citep{arts2018}. By minimizing predictive latency in citation metrics, our system provides a scalable solution for automated high-precision IP valuation.

Our study pioneers a theoretical framework that aligns LLMs' semantic reasoning with revealed economic preferences to establish a new paradigm for real-time and interpretable patent valuation. It has at least threefold research contributions, summarized as follows:

\begin{itemize}
    \item \textbf{Theoretical Contribution via Eco-Semantic Alignment:} We introduce a novel theoretical framework that aligns high-dimensional semantic representations with objective economic renewal signals. By integrating economic signaling theory with natural language processing, this approach reveals how discrete market signals can effectively guide the decoding of complex, unstructured technical specifications, resolving the label ambiguity inherent in unsupervised text analysis. We demonstrate how, by designing specific alignment mechanisms (i.e., Eco-Semantic Alignment), general-purpose AI technology (LLMs) can be successfully adapted to the complex context of IP valuation. This task is highly dependent on professional knowledge, and this adaptation provides a new pathway for realizing intelligent, precise, and reliable IP public services and the strategic management of technological innovation.

    \item \textbf{Methodological Contribution in Reasoning Alignment:} We establish the algorithmic efficacy of the ERA-IT architecture, which synergizes parameter-efficient fine-tuning with domain-specific instruction tuning strategies. We provide empirical evidence that standard pre-trained models are insufficient for the specialized task of IP valuation, demonstrating that our domain-adaptation process aligns the model's semantic processing with the ``Economic Chain-of-Thought'' underlying obfuscated legal-technical texts.

    \item \textbf{Generalizability and Managerial Implications:} We validate the model's capacity to extract universal value signals through extensive empirical analysis of a randomly sampled dataset. Unlike prior studies that were restricted to specific sectors, this work demonstrates the model's applicability across heterogeneous patent texts. By mitigating the systemic latency of bibliometric metrics, the proposed framework empowers stakeholders to make real-time data-driven decisions on the deployment of R\&D resources and the commercialization of technology.
\end{itemize}

Overall, our study provides a novel interdisciplinary lens that bridges information science, information economics, technological innovation management, and explainable AI, thereby enriching the theoretical foundation of context-adaptive innovation. Coherently, it makes contributions through the theoretical conception of Eco-Semantic Alignment and its methodological instantiation, the ERA-IT framework. This synergy offers a new paradigm for the deep adaptation of AI in professional domains, one that is anchored in objective economic reality and realized through an implementable, verifiable algorithmic framework with strong generalization capabilities and rigorous evidentiary standards. As a pioneering study, this work achieves AI innovation in a specific professional context by advancing patent valuation from a field reliant on lagging proxy indicators to a new stage driven by real-time market data and equipped with an interpretable reasoning process.
\section{Related Works}

\subsection{Patent Valuation}
Research on patent valuation has shifted from simple patent counts toward richer indicators that proxy the economic and innovative impact of inventions~\citep{kalip2025,schmitt2025}. Using the American Inventors Protection Act, which mandated earlier publication of most US patent applications, \citet{liu2023} explore a multi-task learning based identification model to identify the high-value patents and the standard-essential patents. \citet{hegde2023} show that earlier disclosure is associated with more and faster follow-on citations, fewer abandonments, and higher R\&D investment, highlighting how publication rules shape both the private and social value of patents. \citet{svensson2022} analyzes four widely used patent value indicators, renewal length, forward citations, family size, and documents that each capture distinct facets of technological and economic value rather than a single underlying measure. Building on this indicator-based tradition, \citet{ponta2021} propose the Innovation Patent Index (IPI), a composite innovation-performance measure that aggregates multiple patent characteristics related to technological quality, market potential, and scope into a firm-level index. Complementing these firm- and patent-level perspectives, \citet{park2023} apply the previously introduced CD index of disruption to large-scale datasets of papers and patents, extend it with normalized variants and alternative measures, and show that science and technology have become less disruptive over time, illustrating the potential of advanced citation-based metrics for characterizing the qualitative role of inventions in the knowledge network. \citet{lee2025} proposes a machine learning framework based on structured bibliometrics and inventor-related indicators. By introducing three underexplored value metrics and applying six algorithms across five distinct strategies to predict the value of renewable energy patents, the findings demonstrate that random forest and artificial neural network models, which integrate multiple indicator categories, achieve the highest accuracy.

However, from the perspective of ex-ante patent valuation, these approaches remain largely ex-post and rely on outcome variables that materialize only after a substantial time lag~\citep{fu2025,liu2024a}. Renewal behavior, composite indices such as IPI, and disruption-based citation metrics are typically calibrated on long-term renewal patterns, realized commercialization, or accumulated citation structures, and they make limited use of the full semantic content of patent texts at the level of individual inventions~\citep{svensson2022,ponta2021,park2023}. Moreover, these indicators generally summarize value at the firm or field level and do not provide explicit, stepwise economic rationales for why a particular patent should be assessed as more or less valuable at the time of filing. Because of these limitations, existing patent-valuation approaches remain constrained in their ability to provide economically grounded and explainable ex-ante value assessments for newly filed patents. These limitations underscore the need for ex-ante patent valuation frameworks that integrate semantic text analysis with interpretable reasoning, such as ERA-IT.

\subsection{Semantic Analysis for IP Management}
Recent work in IP management increasingly relies on semantic analysis of patent and related technical text~\citep{jiang2025}. \citet{ali2024} review patent prior-art retrieval, tracing the field’s progression from basic Boolean and keyword search and classic information
retrieval models toward semantic retrieval enabled by NLP and deep learning. \citet{xu2025} extend this line of work to industry–academia collaboration by proposing a semantic retrieval approach that uses text embeddings to analyze and rank collaborative research outputs. \citet{yun2022} propose a method for mining technology trends based on patent semantic analysis. Through a four-stage process involving the extraction of technical purposes and effects, clustering-based lexicon construction, mining of evolutionary patterns, and forecasting future directions by referencing models from other fields, it supports enterprises in identifying new technology opportunities. At the model level, \citet{chikkamath2022} construct large-scale USPTO subclass datasets and investigate how a BERT for Patents encoder and strong non-neural baselines use claims and abstracts for multi-label patent classification. Complementary research on structured representations decomposes Chinese patent claims into function, structure, and purpose segments and applies pre-trained transformer-based models to classify them into technical categories, thereby improving classification accuracy and supporting retrieval of patents in similar technical domains~\citep{li2025}. In parallel, \citet{zini2022} survey explainability techniques for deep NLP models, cataloguing model-agnostic and model-specific methods that interpret embeddings, internal representations, and output decisions for text-based tasks. \citet{yoshikawa2025} enhance automatic patent classification by leveraging GPT-3.5-turbo to generate summaries from patent claims and detailed descriptions, which are then used to fine-tune RoBERTa (for English) and XLM-RoBERTa (for Japanese) models.

Despite these advances, existing semantic approaches are optimised for retrieval, classification, or document alignment rather than for transparent ex-ante economic valuation of patents. The patent retrieval survey focuses on relevance and recall, leaving the relationship between semantic similarity and economic value largely implicit~\citep{ali2024}. Semantic matching systems for industry–academia collaboration operate in high-dimensional embedding spaces but do not attempt to infer monetary or strategic value or to incorporate ground-truth financial signals~\citep{xu2025}. Models such as the BERT for Patents and the structured claim classifiers in the Chinese patent study are trained to predict technology codes, not renewal-based value tiers, and typically output numerical labels without explicit economic rationales~\citep{chikkamath2022,li2025,abbasiantaeb2025}. Even where explainability is the primary topic, current surveys emphasise post-hoc interpretation tools for generic NLP tasks and note that rigorous evaluation in high-stakes legal and economic contexts remains limited~\citep{zini2022}. In contrast, ERA-IT explicitly fuses semantic reasoning with revealed preference renewal data and uses instruction-tuned chain-of-thought generation to map claim language to value tiers, positioning ERA-IT as a bridge between semantic IP analytics and transparent economic valuation while addressing both the opacity of black-box language models and the latency of traditional citation-based indicators.

\subsection{Generative AI for IP Management}
Recent work on instruction-tuned generative models provides the technical basis for structured reasoning in ERA-IT. Instruction tuning surveys indicate that LLMs can be further trained on collections of instruction-response pairs to improve controllability and domain adaptation while systematising common pitfalls of instruction-driven alignment~\citep{zhang2023}. Chain-of-Thought prompting studies demonstrate that demonstrations with explicit intermediate steps can markedly improve multi-step reasoning, and that relevance and step ordering matter more than strictly correct logical derivations~\citep{wang2023cot}. Research on self-explanations finds that model-generated rationales may appear plausible yet diverge from traditional attribution methods, motivating explicit checks of explanation faithfulness~\citep{huang2023selfexplain}. In high-stakes legal settings, participatory design with practising lawyers argues that LLMs should primarily function as decision-support tools within explicit policy frameworks rather than as autonomous legal advisers~\citep{cheong2024law}. \citet{xiong2025} propose a large language model-assisted active learning framework designed for scalable multi-label patent classification, validated on a dataset of 100,000 drone-related patents. By integrating GPT-4’s uncertainty estimation with diversity-aware sampling, this approach significantly reduces the need for manual annotation while maintaining strong classification performance and improving key evaluation metrics. Within the IP domain, domain-specific frameworks guide models along patent-relevant dimensions~\citep{shomee2025}. For example, PatentMind uses a multi-aspect reasoning graph to decompose and score patent similarity across technical, legal, and contextual dimensions~\citep{yoo2025patentmind}; PatentScore defines a multi-dimensional rubric for evaluating LLM-generated claims across structural, semantic, and legal aspects~\citep{yoo2025patentscore}; and Self Filtered Distillation uses LLM-generated labels and rationales to compute trust scores, thereby filtering or reweighting training examples for patent classification~\citep{yoo2025sfd}.

However, from the perspective of ex-ante patent valuation, these techniques remain only loosely connected to economic ground truth. Instruction tuning and Chain-of-Thought prompting are typically optimised on general reasoning benchmarks and do not directly constrain rationales or predictions with observable economic outcomes like renewal behaviour, so their explanations are at best indirectly related to value signals~\citep{zhang2023,wang2023cot}. Studies of self-explanations further show that natural language rationales can appear faithful under coarse metrics while assigning very different importance patterns, which complicates their use as evidence for economic reasoning~\citep{huang2023selfexplain}. In the patent domain, PatentMind and PatentScore align model behaviour with expert assessments of similarity and claim quality, and Self Filtered Distillation improves the reliability of patent classifiers using trust indicators derived from model outputs, yet none of these frameworks explicitly tie reasoning to renewal trajectories or other revealed preference signals~\citep{yoo2025patentmind,yoo2025patentscore,yoo2025sfd}. Therefore, ERA-IT couples patent renewal history, treated as revealed economic preference, with a four-step Economic Chain-of-Thought schema covering technical strength, market context, legal robustness, and value synthesis, and uses teacher-guided data synthesis together with parameter-efficient adaptation to align LLM reasoning with these economic signals and to turn generated explanations into actionable features for ex-ante patent valuation.

\section{Research Objectives}

We explore three critical research questions underexplored in prior studies:
\begin{itemize}
    \item \textbf{RQ1 (Efficacy):} Does ERA-IT outperform state-of-the-art discriminative baselines and LLMs in predicting patent value?
    \item \textbf{RQ2 (Mechanism):} How does the explicit generation of economic reasoning contribute to the model's predictive performance?
    \item \textbf{RQ3 (Generalizability):} Is the framework robust across heterogeneous technological domains (IPC sections) and low-resource settings?
\end{itemize}

Based on the proposed framework and the defined research inquiries, this study pursues three specific objectives: (1) To empirically substantiate the predictive efficacy of the ERA-IT framework against state-of-the-art discriminative baselines and zero-shot generative LLMs for patent value prediction, using patent renewal history as the ground-truth signal. This objective directly tests the core claim that aligning LLM reasoning with revealed economic preferences, a process known as Eco-Semantic Alignment, yields superior and timely performance metrics. (2) To deconstruct and explain the contributory mechanism of explicit economic rationale generation. Through ablation studies and comparative analysis, this objective aims to isolate and quantify the impact of the Economic Chain-of-Thought module, thereby elucidating how logically grounded rationales enhance model fidelity and serve as a transparent cognitive scaffold for decision-making. (3) To rigorously verify the robustness and generalizability of the framework across heterogeneous technological domains classified by IPC sections and under low-resource conditions. This objective evaluates the practical utility and scalability of ERA-IT, ensuring its reliability beyond the training distribution and confirming its applicability to diverse innovation management contexts.

\section{Dataset}

\subsection{Sampling Strategy and Longitudinal Alignment}
To establish a valuation framework with high external validity, it is critical to mitigate the selection bias often present in domain-specific studies. Unlike prior works restricted to high-velocity sectors, such as semiconductors or biotech, we aimed to capture generalized linguistic patterns of economic value applicable across the entire patent ecosystem.

Accordingly, our data collection followed a rigorous, randomized, stratified sampling strategy. We targeted historical patent filings within the EPO database, specifically defining the observation cohort to the 2010--2015 period. This timeframe ensures an \textit{ex post} observation horizon of at least eight years, allowing for a complete retrospective analysis of economic survival trajectories.

The final analytical corpus comprises 10,000 patent documents uniformly sampled across all primary IPC sections (A--H). By enforcing this heterogeneity, we ensure that the model learns to extract universal economic signals rather than overfitting to the technical terminologies of specific industries. Table~\ref{tab:dataset_examples} presents representative examples derived from the constructed ground-truth dataset.

\begin{table}[ht]
    \centering
    \caption{Representative examples of the constructed ground-truth dataset. The corpus consists of 10,000 patents filed between 2010 and 2015, randomly sampled across diverse IPC sections (A--H). This randomized stratification ensures that the proposed framework captures universal value signals across heterogeneous technological landscapes.}
    \label{tab:dataset_examples}
    \resizebox{\textwidth}{!}{%
    \begin{tabular}{@{}llclcl@{}}
    \toprule
    ID & IPC Section & Filing Year & Subject Matter (Summary) & Renewals & Assigned Label \\ \midrule
    Pat. 1 & A (Human Necessities) & 2011 & Manual coffee brewing mechanism & 2 yrs & Class 1 (Low) \\
    Pat. 2 & B (Transporting) & 2013 & Vehicle side mirror housing & 5 yrs & Class 2 (Mid) \\
    Pat. 3 & G (Physics/Computing) & 2014 & Neural network quantization & 9 yrs & Class 3 (High) \\
    Pat. 4 & C (Chemistry) & 2012 & CRISPR-Cas9 gene editing vector & 11 yrs & Class 3 (High) \\
    Pat. 5 & H (Electricity) & 2010 & Legacy 3G communication protocol & 3 yrs & Class 1 (Low) \\ \bottomrule
    \end{tabular}%
    }
\end{table}

\subsection{Operationalizing Value via Revealed Preferences}
To construct objective ground-truth labels for supervised learning, we operationalized patent renewal records as a proxy for private economic value, grounded in the theory of revealed preferences~\citep{danish2020,harhoff2003,jou2018,porter2023}. Theoretically, the payment of annual renewal fees constitutes a ``costly signal.'' It reflects the patent holder's internal assessment that the expected future returns (or option value) of the asset exceed the cumulative and monotonically increasing maintenance expenses~\citep{pakes1986}.

In the context of information systems, this financial decision serves as a definitive binary signal at each time step, encoding the latent utility of the invention free from the noise of ``cheap talk.'' Unlike citation metrics, which suffer from significant lag in accumulation and potential strategic manipulation, renewal behavior provides a real-time, forward-looking indicator of value retention. This enables converting discrete financial transactions into a robust target variable for algorithmic prediction~\citep{kumar2024}.

\subsection{Label Discretization and Value Stratification}
To convert the continuous-time series of renewal events into a format suitable for supervised learning, we stratified the corpus into three discrete value tiers. This tri-modal discretization is not arbitrary; it is empirically grounded in the marginal cost-benefit analysis of the EPO's progressive fee structure and the typical hazard rate distribution observed in our dataset~\citep{danish2020,jou2018,kumar2024,porter2023}. By mapping renewal counts to specific economic regimes, we capture the latent utility of each asset as follows:

\begin{itemize}
    \item \textbf{Class 1 (Low Value - Early Screening):} Patents renewed one to three times. This tier represents the lower quartile of the value distribution. In this phase, maintenance fees are relatively nominal; thus, a decision to relinquish rights signals an early ``fail-fast'' realization. Economically, this suggests the patent holder identified a lack of commercial viability or technical obsolescence almost immediately following the grant, opting to avoid further sunk costs.

    \item \textbf{Class 2 (Medium Value - Standard Cycle):} Patents renewed four to six times. These patents represent the corpus's median behavioral cluster. They demonstrate sufficient utility to justify maintenance during the primary commercialization window. However, they are abandoned before reaching the seventh year, a critical inflection point at which EPO renewal fees begin to escalate sharply. This reflects ``standard'' assets that provide moderate protection but lack the strategic necessity to warrant high-cost long-term retention.

    \item \textbf{Class 3 (High Value - Strategic Assets):} Patents renewed seven or more times. This category captures the long tail of high-value IP and top-tier assets. Given the exponentially increasing cost of maintaining EPO patents as they approach their second decade, continued renewal beyond the seventh year signals a robust trajectory toward long-term retention. These documents represent technologies with sustained competitive advantages, high licensing potential, or significant strategic blocking value that far outweighs the cumulative financial burden of maintenance.
\end{itemize}

The resulting distribution provides a balanced target variable that reflects the revealed economic conviction of the patent holders, ensuring the model learns to distinguish between transient technical disclosures and sustained commercial assets.
\section{Methodology: ERA-IT Framework}
\label{sec:methodology}

\begin{figure}[ht]
    \centering
    \includegraphics[width=\textwidth]{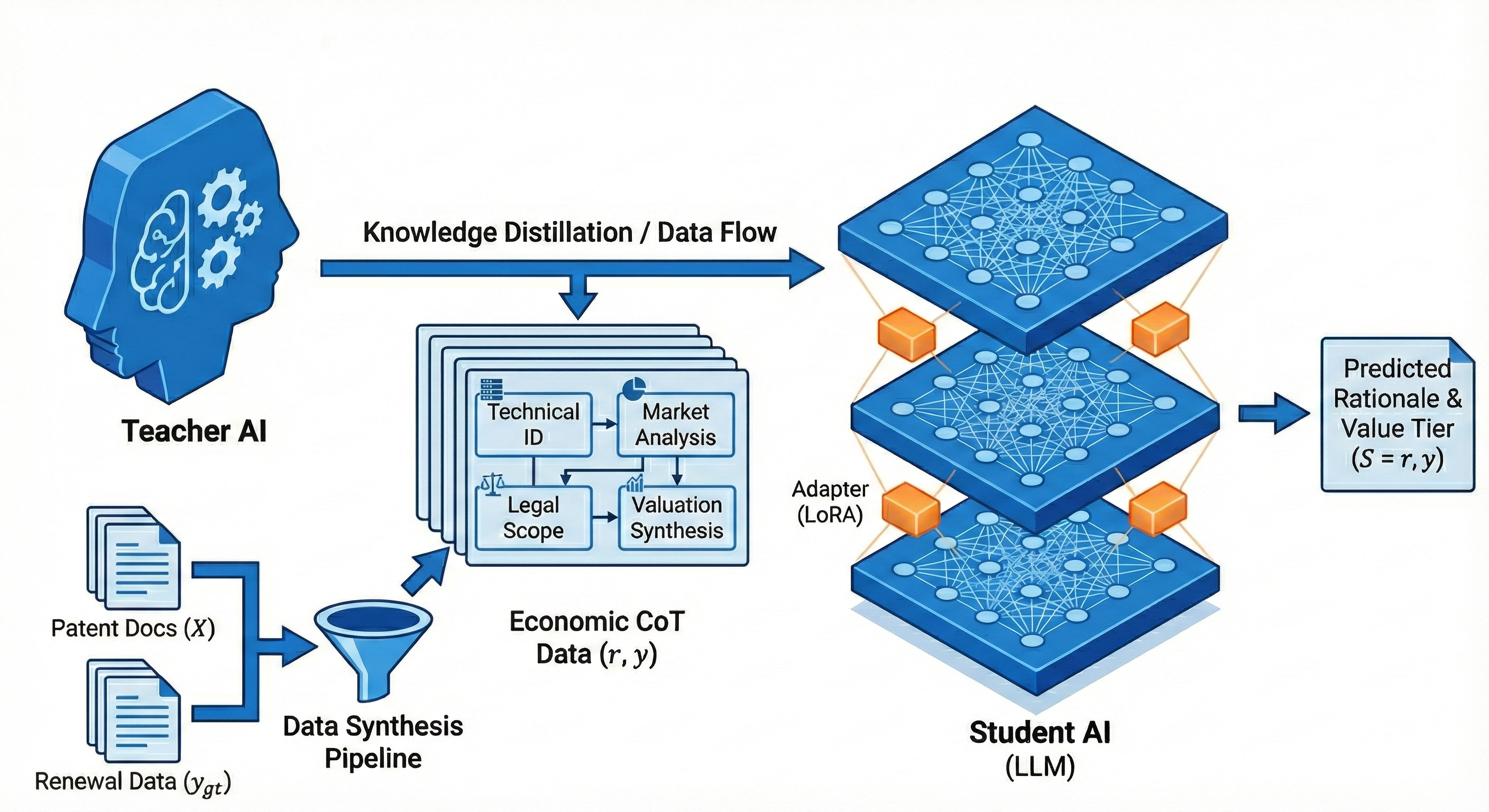}
    \caption{The schematic overview of the ERA-IT framework. The pipeline integrates teacher-guided data synthesis with parameter-efficient instruction tuning to align the model with economic reasoning.}
    \label{fig:framework_pipeline}
\end{figure}

In this section, we propose the ERA-IT framework. Unlike traditional discriminative approaches that treat patent valuation as a black-box classification task, our framework adopts a generative paradigm to explicitly model the causal reasoning process underlying economic valuation. By aligning LLMs' semantic processing capabilities with objective market signals, we aim to bridge the gap between high-dimensional technical specifications and their realized private value. Figure~\ref{fig:framework_pipeline} illustrates the overall pipeline of the proposed architecture.

\subsection{Problem Formulation: From Classification to Causal Reasoning}

Traditional patent valuation is typically formulated as maximizing the conditional probability $P(y|X)$, where $y \in \mathcal{Y}$ represents the discrete value tier given the patent document $X$. However, this formulation suffers from interpretative opacity; it fails to explicate ``why'' a patent holds value, often relying on spurious correlations within the text.

We redefine this task as a Conditional Reasoning Generation problem, aiming to model the Economic Chain-of-Thought. Our objective is to generate a sequence $S = (r, y)$ that includes $r$, the logical rationale justifying the valuation (e.g., interpretation of exclusionary scope, analysis of barriers to entry), followed by the prediction $y$. Consequently, the model parameters $\theta$ are optimized to maximize the following likelihood:

\begin{equation}
    \theta^* = \operatorname*{argmax}_\theta \sum_{(X, r, y) \in \mathcal{D}} \log P_\theta(r, y | X)
\end{equation}
where $\mathcal{D}$ represents the instruction tuning dataset constructed to capture the causal link between technical disclosure and economic survival.

\subsection{Reverse-Engineering the Economic Chain-of-Thought}
\label{sec:data_construction}

A critical bottleneck in supervised learning for valuation is the scarcity of high-quality reasoning data ($r$). To address this, we propose a semi-automated pipeline that uses patent renewal data as objective ground truth to reverse-engineer the latent economic logic.

\subsubsection{Structural Schema for Cognitive Alignment}
To ensure the generated reasoning aligns with domain experts' cognitive processes, we define a four-step reasoning schema. This schema forces the model to sequentially evaluate the structural components of patent value before reaching a conclusion:

\begin{itemize}
    \item \textbf{Technical Identification ($\phi_{tech}$):} The model distills the technical essence of the invention, filtering out obfuscated legalese to identify the core novelty.
    \item \textbf{Market Application Analysis ($\phi_{market}$):} The model infers the potential industrial application and market size, estimating the demand-side dynamics.
    \item \textbf{Exclusionary Scope Interpretation ($\phi_{legal}$):} The model evaluates the appropriability of the invention by analyzing the breadth of independent claims and the difficulty of design-around by competitors.
    \item \textbf{Valuation Synthesis ($\phi_{value}$):} Synthesizing the technical, market, and legal factors, the model establishes a causal link to the realized renewal duration.
\end{itemize}

\subsubsection{Teacher-Guided Data Synthesis}
\begin{figure}[ht]
    \centering
    \includegraphics[width=0.8\textwidth]{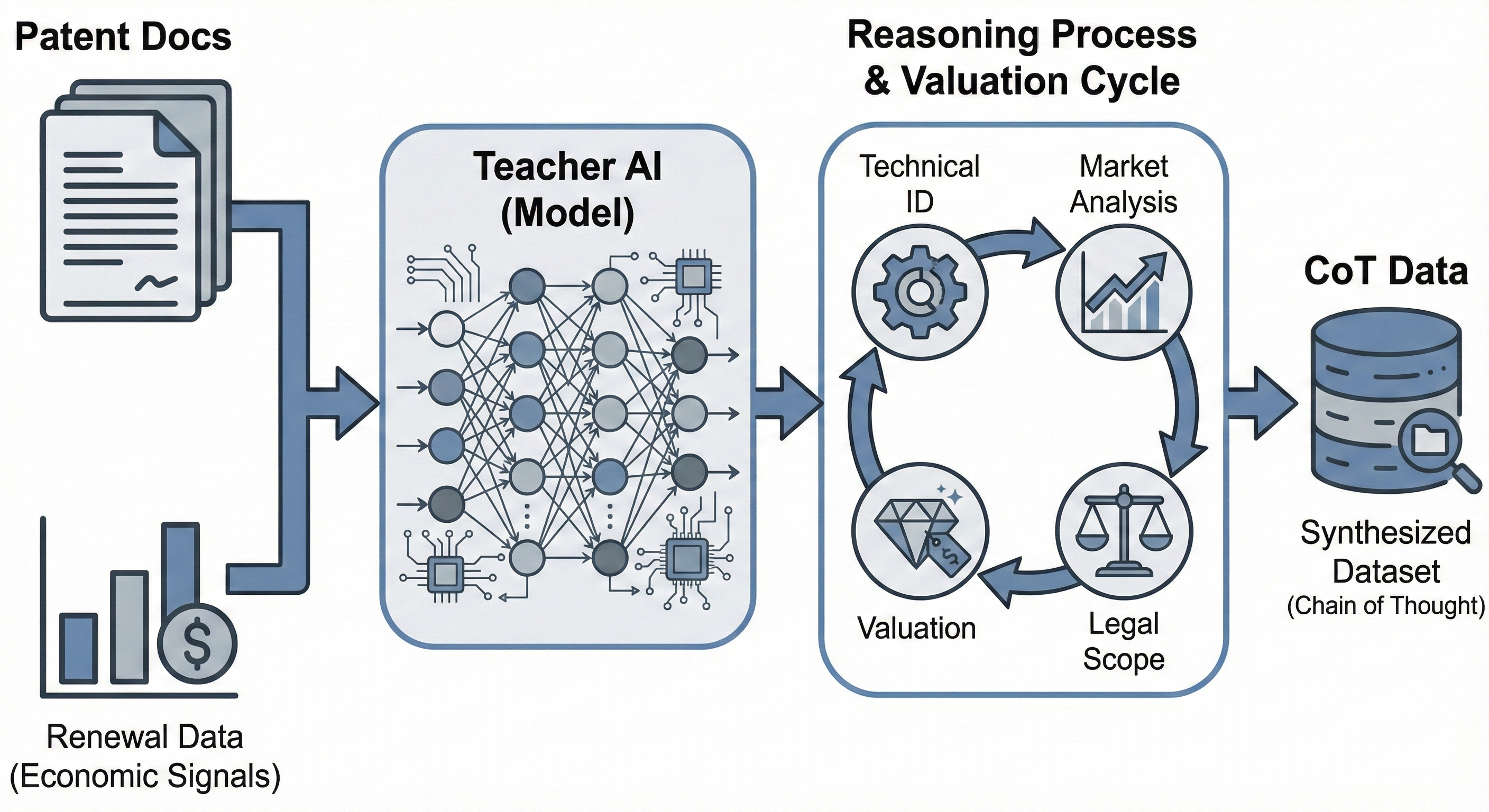}
    \caption{The data synthesis process utilizing a Teacher Model to reverse-engineer economic reasoning based on ground-truth renewal data.}
    \label{fig:data_synthesis}
\end{figure}

For the training corpus $X_{train}$ with actual renewal history, we leverage a high-performance Teacher Model (specifically, GPT-4) to synthesize the reasoning data $\hat{r}$. Figure~\ref{fig:data_synthesis} illustrates the process of using the teacher model to conduct data synthesis. We employ a \textit{Label-Conditioned Generation} strategy by providing the ground-truth label $y_{gt}$ as an anchor in the prompt. This induces the model to construct an \textit{ex post} rationalization based on the actual economic outcome. The generated $\hat{r}$ serves as the pseudo-cognitive path that logically bridges the document $X$ and the value $y$. The final dataset is curated as $\mathcal{D}_{CoT} = \{(X_i, r_i, y_i)\}_{i=1}^N$.

\subsection{Domain Adaptation via Instruction Tuning}

We fine-tune a Student LLM (e.g., Llama-3-8B) using the constructed $\mathcal{D}_{CoT}$. To effectively inject domain-specific economic logic while preserving the model's general linguistic capabilities, we adopt a parameter-efficient training strategy using LoRA (Low-Rank Adaptation).

\subsubsection{Instruction-Formatted Input Representation}
The raw patent document $X$ is transformed into a structured prompt in a hierarchical tag format. The input sequence $I$ is defined as follows:

\begin{equation}
    I = \text{[SYS]} \oplus \mathcal{I}_{task} \oplus \text{[META]} \oplus M \oplus \text{[DOC]} \oplus X \oplus \text{[ANS]}
\end{equation}
where $\mathcal{I}_{task}$ denotes the role-playing instruction (e.g., ``You are an expert IP analyst''), forcing the model to adopt the persona of a valuation expert, and $\oplus$ denotes string concatenation.

\subsubsection{LoRA-based Adaptation}
We freeze the pre-trained weight matrix $W_0 \in \mathbb{R}^{d \times k}$ and update the learning parameters by introducing low-rank approximation matrices $A \in \mathbb{R}^{d \times r'}$ and $B \in \mathbb{R}^{r' \times k}$, where the rank is $r' \ll \min(d, k)$. The weight update is formulated as:

\begin{equation}
    W = W_0 + \Delta W = W_0 + \frac{\alpha}{r'} B A
\end{equation}
where $\alpha$ is a scaling factor. This approach allows for securing domain alignment by training less than 0.1\% of the parameters, preventing catastrophic forgetting of general reasoning capabilities.

\subsection{Optimization Objective with Instruction Masking}

Model training follows the Causal Language Modeling loss function. Crucially, we apply \textit{Instruction Masking} to prevent the model from merely memorizing the input patent text $X$.

Let $T = [t_1, \dots, t_L]$ be the tokenized sequence of the formatted input $I$, and $L_{input}$ be the length of the prompt segment (instruction and document). The optimization loss function $\mathcal{L}$ is defined as:

\begin{equation}
    \mathcal{L}(\theta) = - \sum_{i=L_{input}+1}^{L} \log P_\theta(t_i | t_1, \dots, t_{i-1}; X)
\end{equation}

By calculating the loss exclusively on the economic rationale and value tier segments ($r$ and $y$), we force the model to focus on learning the conditional probability distribution that yields a logical valuation consistent with the specification. This distinguishes our framework from traditional approaches that predict outcomes without an explicit reasoning process.

\subsection{Inference and Value Prediction}

During the inference phase, the model is provided with the input prompt $I_{test}$ containing only the instruction and the patent document. The model generates the token sequence autoregressively:

\begin{equation}
    S_{gen} = \operatorname*{argmax}_{S} P_\theta(S | I_{test})
\end{equation}

Since the model is trained to generate the output in a structured format, we employ a deterministic parsing function $\psi(\cdot)$ to extract the predicted value tier $\hat{y}$. This end-to-end generation process allows stakeholders to obtain not only the valuation score but also the interpretative transparency required for cognitive scaffolding in decision-making.
\section{Experiments}
\label{sec:experiments}

To rigorously validate the proposed ERA-IT framework, we designed comprehensive experiments guided by the three research questions proposed in Section 1.

\subsection{Experimental Setup}

\subsubsection{Datasets and Evaluation Metrics}
We utilized the dataset of 10,000 EPO patents constructed in Section~\ref{sec:data_construction}. The dataset was stratified and randomly split into training (8,000), validation (1,000), and testing (1,000) sets.
Given the potential class imbalance in real-world scenarios, we report not only Accuracy but also the Macro-F1 score and the Matthews Correlation Coefficient (MCC), which is considered a more robust metric for imbalanced classification tasks.

\subsubsection{Baselines}
We compared ERA-IT against three categories of strong baselines:
\begin{itemize}
    \item \textbf{Traditional machine learning:} \textit{TF-IDF+SVM} and \textit{TF-IDF+Random Forest} serve as fundamental benchmarks for text classification.
    \item \textbf{Discriminative pre-trained language models:} We included \textit{BERT-Large}, \textit{SciBERT} (trained on scientific text), and \textit{Longformer} (optimized for long documents) to evaluate the limitations of encoder-only models.
    \item \textbf{Generative LLMs:} We evaluated \textit{Deepseek-r1-8B}, \textit{Llama-3-8B}, \textit{Qwen-3-8B} (base model) in a 5-shot setting and \textit{GPT-5-mini} (zero-shot) to measure the gap between general reasoning and domain-specific instruction tuning.
\end{itemize}

\subsubsection{Implementation Details}
The Student AI (Llama-3-8B) was fine-tuned using LoRA with rank $r=16$, $\alpha=32$, and dropout $0.05$. We used the AdamW optimizer with a cosine learning rate schedule (initial $lr=2e-4$) and a warm-up ratio of 0.03. All experiments were conducted on a single NVIDIA H100 GPU.

\subsection{Performance Analysis (RQ1)}
The ERA-IT framework achieves superior performance across all metrics, as shown in Table 2. Notably, it outperforms the traditional Random Forest model based on TF-IDF features by 36.3\% in accuracy, 33.1\% in Macro-F1, and 88.1\% in MCC. It outperforms the discriminative pre-trained language model Longformer by 11.5\% in accuracy, 8.7\% in Macro-F1, and 27.4\% in MCC, suggesting that generative reasoning is more effective for valuation than merely processing long contexts. Furthermore, it outperforms the zero-shot performance of the generative LLM GPT-5-mini by 9.6\% in accuracy, 8.3\% in Macro-F1, and 21.5\% in MCC, demonstrating that parameter-efficient fine-tuning with economic signals yields better domain alignment than the general reasoning capability of closed-source giant models.

\begin{table}[ht]
    \centering
    \caption{Comparative performance on the test set.† indicates statistical significance (p < 0.01) compared to the best baseline.}
    \label{tab:main_results}
    \resizebox{\textwidth}{!}{%
    \begin{tabular}{@{}lcccc@{}}
    \toprule
    \textbf{Model Type} & \textbf{Model} & \textbf{Accuracy} & \textbf{Macro-F1} & \textbf{MCC} \\ \midrule
    \multirow{2}{*}{Traditional} & TF-IDF + SVM & 58.4 & 56.2 & 0.38 \\
     & TF-IDF + Random Forest & 61.2 & 59.8 & 0.42 \\ \midrule
    \multirow{3}{*}{Discriminative} & BERT-Large & 69.5 & 68.1 & 0.54 \\
     & SciBERT & 73.2 & 71.8 & 0.59 \\
     & Longformer & 74.8 & 73.2 & 0.62 \\ \midrule
    \multirow{4}{*}{Generative} & Deepseek-r1-8b (5-shot) & 64.2 & 61.8 & 0.51 \\
     & Llama-3-8B (5-shot) & 68.9 & 67.4 & 0.52 \\
     & Qwen-3-8B (5-shot) & 69.4 & 69.1 & 0.49 \\
     & GPT-5-mini (Zero-shot) & 76.1 & 73.5 & 0.65 \\ \midrule
    \textbf{Ours} & \textbf{ERA-IT (Full)} & \textbf{83.4}$^\dagger$ & \textbf{79.6}$^\dagger$ & \textbf{0.79}$^\dagger$ \\ \bottomrule
    \end{tabular}%
    }
    \end{table}

\subsection{Ablation and Mechanism Analysis (RQ2)}

To isolate the contribution of the Economic Chain-of-Thought reasoning, we conducted an ablation study, as shown in Table 3. 

\begin{table}[ht]
    \centering
    \caption{Ablation study on model components. Removing each component results in a performance drop, confirming its contribution to the framework.}
    \label{tab:ablation}
    \begin{tabular}{@{}lc@{}}
    \toprule
    \textbf{Configuration} & \textbf{Macro-F1 ($\Delta$)} \\ \midrule
    \textbf{ERA-IT (Full Framework)} & \textbf{79.6} \\
    \quad w/o Rationale Generation ($r$) & 74.5 (-5.1) \\
    \quad w/o Instruction Masking & 77.2 (-2.4) \\
    \quad w/o Hierarchical Tags (Flat Text) & 78.4 (-1.2) \\ \bottomrule
    \end{tabular}
\end{table}

The most significant performance drop (-6.4\%) occurs when the rationale generation ($r$) is removed. This confirms that the generated reasoning acts as a ``cognitive scaffold,'' forcing the model to attend to relevant technical and legal details before making a prediction. Removing instruction masking also degrades performance, indicating that focusing the loss solely on the output helps the model learn the generation task more efficiently.

\subsection{Robustness and Generalizability (RQ3)}

\subsubsection{Performance Across Technology Domains}
To verify the generalizability claimed in the Introduction, we analyzed the model's accuracy across different IPC sections. As shown in Figure 5, the model maintains high accuracy ($>80\%$) in Physics (G) and Electricity (H), and shows robust performance even in Chemistry (C) and Mechanics (B), confirming its applicability across heterogeneous patent landscapes.

\begin{figure}[h!]
    \centering
    \includegraphics[width=0.8\textwidth]{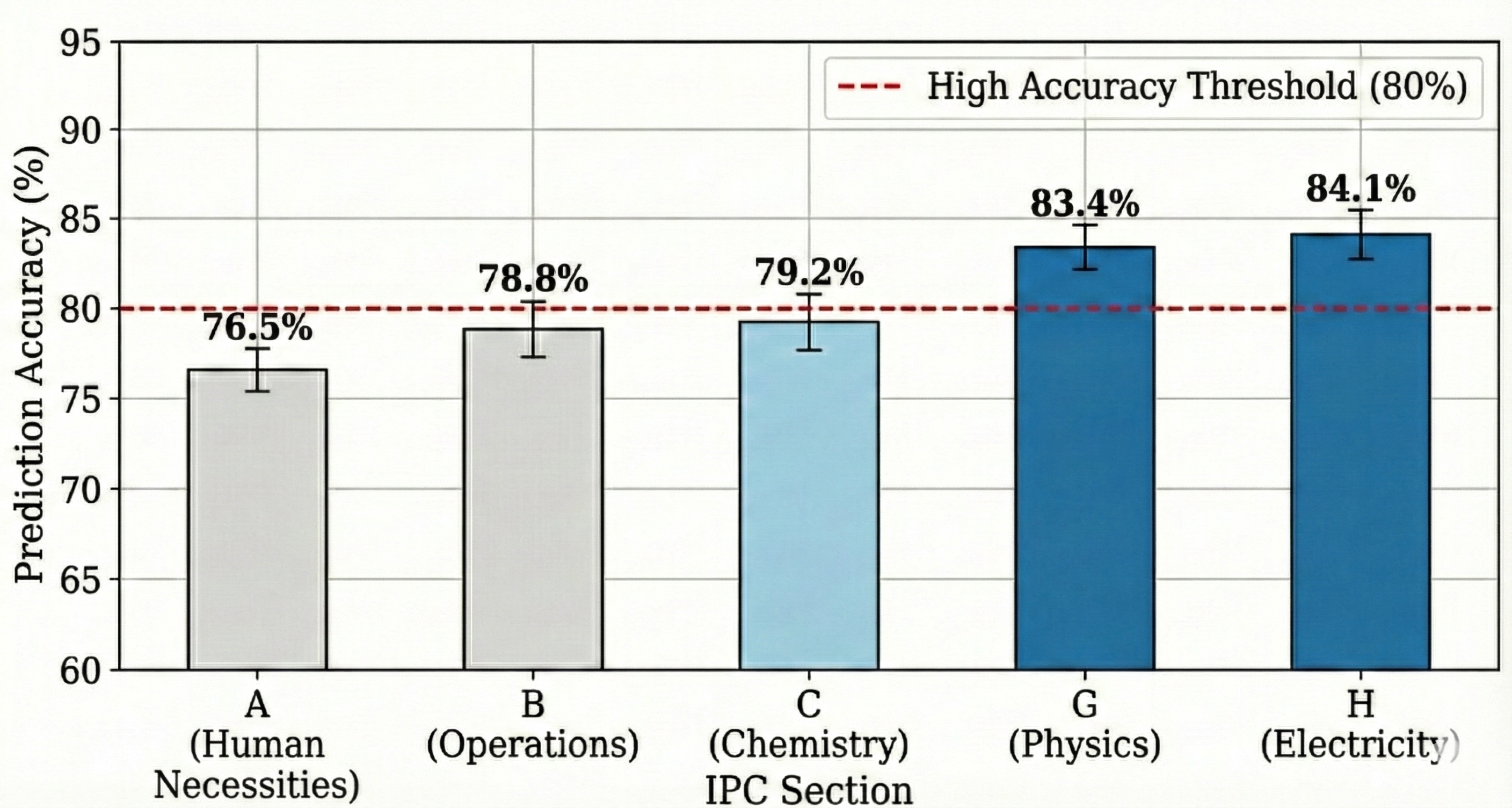}
    \caption{Prediction accuracy across major IPC sections. The model demonstrates consistent performance across technological domains.}
    \label{fig:ipc_performance}
\end{figure}

\subsubsection{Sample Efficiency}
We further evaluated the model's performance under low-resource settings by varying the training data size from 10\% to 100\%. ERA-IT achieves 90\% of its peak performance with only 40\% of the data, suggesting that the Economic Chain-of-Thought data provides dense supervisory signals, enabling efficient learning even with limited labeled data.

\subsection{Qualitative Case Study}
To demonstrate the interpretability of ERA-IT, we present a representative inference example in Table~\ref{tab:case_study}. Unlike baseline models that only output a probability score, our framework explicitly identifies the ``broad range of particle sizes'' (0.01-10 $\mu$m) in the input claim and logically connects it to a ``strong exclusionary scope.'' This narrative validates that the model has successfully internalized the economic reasoning patterns from the renewal data.

\begin{table}[h!]
    \centering
    \caption{A generated inference example for a Class 3 patent (Solid-state Electrolyte).}
    \label{tab:case_study}
    \begin{tabular}{|p{0.95\textwidth}|}
    \hline
    \textbf{Input Claim Fragment:} \\
    \textit{"...comprising a solid-state electrolyte layer having inorganic particles with an average diameter of \textbf{0.01 to 10 $\mu$m} distributed therein to prevent dendrite growth..."} \\
    \hline
    \textbf{Model Generated Rationale ($r$):} \\
    \small
    1. \textbf{Technical Identification:} The invention specifies a solid-state electrolyte composition, focusing on particle distribution for safety. \\
    2. \textbf{Legal Scope Analysis:} The claimed particle size range (0.01-10 $\mu$m) is designed to be \textbf{intentionally broad}. This creates a strong exclusionary scope, making it difficult for competitors to design around using similar ceramic materials without infringing. \\
    3. \textbf{Market Viability:} This feature is critical for EV battery safety, implying high commercial demand. \\
    \hline
    \textbf{Predicted Value Tier ($y$):} \textbf{Class 3 (High Value)} \\
    \hline
    \end{tabular}
\end{table}
\section{Theoretical, Methodological, and Practical Implications}
As noted in Section 2, existing patent-valuation approaches are constrained in providing economically grounded, explainable ex-ante assessments. Their limitations are threefold. First, they are largely ex-post, relying on outcome variables that materialize with substantial lag~\citep{fu2025,liu2024a}. Second, metrics based on renewal behavior, composite indices, or citation structures are calibrated on long-term patterns and make limited use of the semantic content within individual patent texts~\citep{svensson2022,ponta2021,park2023}. Third, these indicators aggregate value at the firm or field level and fail to provide stepwise economic rationales for a patent's value at filing. To overcome these shortcomings, our study proposes a novel framework, ERA-IT, which integrates deep semantic analysis with interpretable reasoning for true ex-ante valuation. Compared with existing research, our study has significant theoretical, methodological, and practical implications.

\subsection{Theoretical Implications}
This research offers substantial theoretical contributions to the literature on information systems, AI innovation, and IP Management by establishing Eco-Semantic Alignment as a new paradigm connecting economic reality with AI semantic reasoning.

First, we introduce revealed economic preference as a novel supervision benchmark for AI domain adaptation. We elucidate that traditional citation-based bibliometric indicators suffer from systemic latency and fail to provide effective signals for real-time valuation. In contrast, the patent renewal decision is a market action taken by the right holder based on private information, directly reflecting the expected economic value of the patent. The core theoretical contribution of this research lies in systematically demonstrating that patent renewal history, a discrete and observable market behavior, serves as an objective and timely ground truth for Large Language Models to decode the semantics of patent texts. This finding challenges the prevailing reliance on bibliometrics and shows that financial maintenance decisions offer a more immediate, less noisy proxy for technological value. By directly mapping linguistic patterns to economic survival, our work transcends the label ambiguity inherent in traditional unsupervised text analysis. It provides a stable anchor in economic logic for decoding complex technical documents, thereby expanding the theoretical understanding of how value is encoded within technical documentation. By algorithmically merging high-dimensional and unstructured technical text with low-dimensional yet information-rich economic behavior data, the ERA-IT framework theoretically empowers firms to conduct context-adaptive innovation in an increasingly digital and dynamic environment.

Second, we define the Economic Chain-of-Thought as the mechanism to bridge the cognitive gap between semantic representation and market value, extending the application of Generative AI from simple text summarization to Generative Reasoning in Economic Contexts. Our framework transcends the black-box paradigm of simply mapping input text to output value. Its key theoretical innovation is proposing the concept of the Economic Chain-of-Thought as an explainable output that the model must explicitly generate. This concept theoretically aligns the semantic reasoning process of AI with the economic logic of human decision-makers. It requires the model not only to predict an outcome but also to reverse-engineer the chain of reasoning that supports a specific economic decision, such as the choice to renew or abandon a patent, from obfuscated legal and technical texts. This process builds a transparent cognitive scaffold for AI reasoning and transforms outputs from opaque predictions into rational arguments that decision-makers can review and adopt. By showing that instruction tuning with domain-specific economic logic significantly outperforms standard pre-training, this study establishes a new benchmark for vertical AI applications where domain expertise is scarce, high-dimensional, and requires logical transparency.

Third, we establish a computable theoretical pathway for firms to achieve AI-driven context-adaptive innovation. This study provides a concrete instantiation of this concept by using a domain-specific and objective economic behavior signal, specifically renewal data, as the supervision signal for instruction tuning. This approach compels a general-purpose LLM to adapt its internal semantic representations to the logic of the specific context of IP valuation. This alignment is not superficial keyword matching but rather drives the model's generative reasoning to conform to the behavioral logic of microeconomic agents, thereby enabling cognitive-level adaptation of the AI system from a general-purpose learner to a domain expert.

Overall, our study establishes a new paradigm to resolve a core contradiction in the field: the chasm between technical complexity and economic measurability.

\subsection{Methodological Implications}
This research designs and validates the ERA-IT executable architecture for achieving Eco-Semantic Alignment.

First, we develop a hybrid training strategy that co-optimizes parameter-efficient fine-tuning and domain-specific instruction. Addressing the dual challenges of patent text specialization and the scarcity of high-quality labeled data, the ERA-IT method avoids computationally intensive full-parameter fine-tuning. Its methodological innovation lies in the synergistic integration of parameter-efficient techniques, such as LoRA, with a carefully constructed set of domain-specific instructions that guide economic reasoning. This approach explicitly demonstrates that for tasks requiring deep domain reasoning, such as patent valuation, standard pre-trained models are insufficient. Instead, a strategy combining lightweight parameter adaptation with explicit task instructions can more effectively capture domain-specific logic while mitigating the risk of overfitting.

Second, we provide empirical evidence linking the establishment of the Economic Chain-of-Thought to model performance. The empirical component of this research goes beyond simply demonstrating improved predictive accuracy. Through structured experimental designs, including ablation studies, it provides evidence that the model's significant performance gain is directly attributable to its ability to generate the Economic Chain-of-Thought. This methodologically confirms that the performance gain stems not from memorizing superficial text features but from the model successfully learning and simulating the internal logical chain connecting technical semantics to economic decisions. This sets a new standard for evaluating the effectiveness of AI models in professional domains, where superior performance must be accompanied by an auditable reasoning process aligned with domain logic.

Third, we validate the method's robust cross-technological domain generalization using Economically Universal Signals. Another key methodological contribution is demonstrating the robustness of the framework across heterogeneous IPC sections. This generalizability benefits from the nature of the supervision signal, specifically renewal behavior, which represents a universal economic logic transcending specific technical details. This indicates that anchoring on high-level, cross-domain, universal behavioral signals is an effective methodological strategy for building specialized AI systems capable of handling technological diversity and overcoming data sparsity challenges in niche domains.

\subsection{Managerial and Practical Implications}
This study holds significant managerial and practical value. The proposed framework serves as a transformative decision-support system for R\&D strategists, patent attorneys, and venture investors, specifically addressing the high cost and latency of due diligence.

\begin{itemize}
    \item \textbf{Mitigating Information Asymmetry via Real-time Valuation:} Innovation markets suffer from severe information asymmetry. Our proposed ERA-IT framework addresses this fundamental challenge by leveraging patent renewal history as a revealed preference signal to align LLMs' generative reasoning with market realities. Unlike citation metrics, which require years to accumulate, our model enables stakeholders to assess the potential value of patent applications immediately upon publication. This enables agile portfolio management and timely technology-transfer decisions in high-velocity markets. Firms can conduct real-time, automated economic health scans of their entire patent portfolio and quickly identify dormant-value patents characterized by a strong economic rationale and low maintenance costs, as well as non-performing assets that lack rationale but incur high renewal fees.

    \item \textbf{Cognitive Scaffolding and Human-AI Complementarity:} The ability of ERA-IT to generate explicit rationales acts as a cognitive scaffold for human experts. Instead of replacing human judgment, we propose a workflow based on proposition and verification to ensure high-trust collaboration. First, the AI functions as a high-throughput screening agent that generates value predictions alongside a transparent Economic Chain-of-Thought. Second, the domain expert conducts rapid verification by reviewing the generated rationale to confirm its legal and technical validity, rather than analyzing the raw specification from scratch. This process leverages the complementarity between the computational scalability of AI and the strategic intuition of human experts, thereby minimizing the black-box risk associated with deep learning deployment. The Economic Chain-of-Thought serves as a cognitive augmentation partner, providing a common language for discussion and improving the overall rationality and consistency of team decisions.
\end{itemize}

In summary, by grounding LLMs' semantic reasoning in the objective economic reality of the market, this study offers a powerful new approach to valuing and managing intangible assets. It provides both a telescope for strategic foresight and a microscope for real-time and transparent analysis.

\subsection{Limitations and Future Directions}
Despite its significant contributions, this study is subject to limitations that delineate avenues for future research.

First, while patent renewal data serves as a robust proxy for private value, it is not free from noise. Firms may maintain low-value patents for strategic blocking purposes or abandon high-value patents due to external financial constraints rather than intrinsic asset quality. Future research could refine the ground-truth labels by incorporating auxiliary signals such as litigation records or Standard-Essential Patent declarations to distinguish between intrinsic technological value and strategic defensive value.

Second, the rationales generated by our model rely on the patterns learned from the teacher model. To ensure these rationales fully align with human expert intuition, future studies should conduct rigorous human evaluations involving patent attorneys to quantify the validity and legal accuracy of the generated reasoning.

Finally, our empirical validation was restricted to EPO data and textual modalities. Extending the framework to other jurisdictions, such as the USPTO and China National Intellectual Property Administration, and integrating visual encoders for technical drawings would represent a significant step toward a truly global and multimodal valuation system.
\section{Conclusion}
This paper addresses the persistent challenges of information asymmetry and interpretative opacity in the patent market by proposing the ERA-IT framework. Moving beyond traditional bibliometric approaches that suffer from significant temporal latency, we theoretically redefined patent renewal history as a revealed economic preference. This enabled us to establish a novel supervised learning paradigm that leverages objective market signals to guide the semantic decoding of LLMs. Our empirical analysis, conducted on a dataset of 10,000 EPO patents randomly sampled across diverse technological domains, substantiates the efficacy of this approach. ERA-IT not only achieved significant performance gains over traditional machine learning models, discriminative pre-trained language models, and zero-shot LLMs but also demonstrated robust generalizability across heterogeneous IPC sections. Crucially, it successfully reverse-engineers the Economic Chain-of-Thought, providing logical rationales that bridge the epistemic gap between obfuscated technical specifications and their realized market value.


\bibliographystyle{elsarticle-harv}
\bibliography{meta_data/custom}

@article{hall2005,
  author  = {Hall, Bronwyn H. and Jaffe, Adam and Trajtenberg, Manuel},
  title   = {Market value and patent citations},
  journal = {The RAND Journal of Economics},
  volume  = {36},
  number  = {1},
  pages   = {16--39},
  year    = {2005}
}

@article{reitzig2004,
  author  = {Reitzig, Markus},
  title   = {Strategic management of intellectual property},
  journal = {MIT Sloan Management Review},
  volume  = {45},
  number  = {3},
  pages   = {35--40},
  year    = {2004}
}

@article{ernst2003,
  author  = {Ernst, Holger},
  title   = {Patent information for strategic technology management},
  journal = {World Patent Information},
  volume  = {25},
  number  = {3},
  pages   = {233--242},
  year    = {2003}
}

@article{arts2018,
  author  = {Arts, Sam and Cassiman, Bruno and Gomez, Jian Cheng},
  title   = {Text matching to measure patent similarity},
  journal = {Strategic Management Journal},
  volume  = {39},
  number  = {1},
  pages   = {62--84},
  year    = {2018}
}

@article{lee2020,
  author  = {Lee, Jieh-Sheng and Hsiang, Jieh},
  title   = {Patent classification by fine-tuning BERT language model},
  journal = {World Patent Information},
  volume  = {61},
  pages   = {101965},
  year    = {2020}
}

@article{pakes1986,
  author  = {Pakes, Ariel},
  title   = {Patents as options: Some estimates of the value of holding European patent stocks},
  journal = {Econometrica},
  volume  = {54},
  number  = {4},
  pages   = {755--784},
  year    = {1986}
}

@article{trajtenberg1990,
  author  = {Trajtenberg, Manuel},
  title   = {A penny for your quotes: Patent citations and the value of innovations},
  journal = {The RAND Journal of Economics},
  volume  = {21},
  number  = {1},
  pages   = {172--187},
  year    = {1990}
}

@article{harhoff2003,
  author  = {Harhoff, Dietmar and Scherer, Frederic M. and Vopel, Katrin},
  title   = {Citations, family size, renewals, and the value of patent rights},
  journal = {Research Policy},
  volume  = {32},
  number  = {8},
  pages   = {1343--1363},
  year    = {2003}
}

@incollection{pakes1984,
  author    = {Pakes, Ariel and Schankerman, Mark},
  title     = {The Rate of Obsolescence of Patents, Research Gestation Lags, and the Private Rate of Return to Research Resources},
  booktitle = {R\&D, Patents, and Productivity},
  editor    = {Zvi Griliches},
  year      = {1984},
  pages     = {73--88},
  publisher = {University of Chicago Press},
  address   = {Chicago, IL},
  }

@article{kumar2024,
  author  = {Kumar, A. and Ranjan, P. and Koley, A. and Danish, S.},
  title   = {A new hybrid machine learning model for predicting the renewal life of patents},
  journal = {PLoS ONE},
  volume  = {19},
  number  = {6},
  pages   = {e0306186},
  year    = {2024}
}

@article{hegde2023,
  author  = {Hegde, D. and Herkenhoff, K. and Zhu, C.},
  title   = {Patent publication and innovation},
  journal = {Journal of Political Economy},
  volume  = {131},
  number  = {7},
  pages   = {1845--1903},
  year    = {2023}
}

@article{svensson2022,
  author  = {Svensson, R.},
  title   = {Patent value indicators and technological innovation},
  journal = {Empirical Economics},
  volume  = {62},
  number  = {4},
  pages   = {1715--1742},
  year    = {2022}
}

@article{ponta2021,
  author  = {Ponta, L. and Puliga, G. and Manzini, R.},
  title   = {A measure of innovation performance: The innovation patent index},
  journal = {Management Decision},
  volume  = {59},
  number  = {13},
  pages   = {73--98},
  year    = {2021}
}

@article{park2023,
  author  = {Park, M. and Leahey, E. and Funk, R. J.},
  title   = {Papers and patents are becoming less disruptive over time},
  journal = {Nature},
  volume  = {613},
  number  = {7942},
  pages   = {138--144},
  year    = {2023}
}

@article{ali2024,
  author  = {Ali, A. and Tufail, A. and De Silva, L. C. and Abas, P. E.},
  title   = {Innovating patent retrieval: A comprehensive review of techniques, trends, and challenges in prior art searches},
  journal = {Applied System Innovation},
  volume  = {7},
  number  = {5},
  pages   = {91},
  year    = {2024}
}

@inproceedings{xu2025,
  author    = {Xu, Q. and Ono, H.},
  title     = {Semantic retrieval for analyzing collaborative research in industry--academia collaboration},
  booktitle = {Proceedings of the International Conference on Network-Based Information Systems},
  pages     = {151--162},
  publisher = {Springer Nature Switzerland},
  address   = {Cham},
  year      = {2025}
}

@inproceedings{chikkamath2022,
  author    = {Chikkamath, R. and Parmar, V. R. and Otiefy, Y. and Endres, M.},
  title     = {Patent classification using {BERT}-for-patents on {USPTO}},
  booktitle = {Proceedings of the 2022 5th International Conference on Machine Learning and Natural Language Processing},
  pages     = {20--28},
  year      = {2022}
}

@article{li2025,
  author  = {Li, R. and Yu, W. and Wang, S.},
  title   = {Research on Chinese patent classification based on structured features},
  journal = {Scientific Reports},
  volume  = {15},
  number  = {1},
  pages   = {18036},
  year    = {2025}
}

@article{zini2022,
  author  = {Zini, J. E. and Awad, M.},
  title   = {On the explainability of natural language processing deep models},
  journal = {ACM Computing Surveys},
  volume  = {55},
  number  = {5},
  pages   = {1--31},
  year    = {2022}
}

@article{zhang2023,
  author  = {Zhang, S. and Dong, L. and Li, X. and Zhang, S. and Sun, X. and Wang, S. and Li, J. and Hu, R. and Zhang, T. and Wu, F. and Wang, G.},
  title   = {Instruction tuning for large language models: A survey},
  journal = {arXiv preprint arXiv:2308.10792},
  year    = {2023}

}

@inproceedings{wang2023cot,
  author    = {Wang, B. and Min, S. and Deng, X. and Shen, J. and Wu, Y. and Zettlemoyer, L. and Sun, H.},
  title     = {Towards understanding chain-of-thought prompting: An empirical study of what matters},
  booktitle = {Proceedings of the 61st Annual Meeting of the Association for Computational Linguistics (Volume 1: Long Papers)},
  pages     = {2717--2739},
  year      = {2023}
}

@article{huang2023selfexplain,
  author  = {Huang, S. and Mamidanna, S. and Jangam, S. and Zhou, Y. and Gilpin, L. H.},
  title   = {Can large language models explain themselves? {A} study of {LLM}-generated self-explanations},
  journal = {arXiv preprint arXiv:2310.11207},
  year    = {2023}
}

@inproceedings{cheong2024law,
  author    = {Cheong, I. and Xia, K. and Feng, K. K. and Chen, Q. Z. and Zhang, A. X.},
  title     = {{(A)} {I} am not a lawyer, but...: engaging legal experts towards responsible {LLM} policies for legal advice},
  booktitle = {Proceedings of the 2024 ACM Conference on Fairness, Accountability, and Transparency},
  pages     = {2454--2469},
  year      = {2024}
}

@article{yoo2025patentmind,
  author  = {Yoo, Y. and Xu, Q. and Cao, L.},
  title   = {{PatentMind}: A multi-aspect reasoning graph for patent similarity evaluation},
  journal = {arXiv preprint arXiv:2505.19347},
  year    = {2025}
}

@article{yoo2025patentscore,
  author  = {Yoo, Y. and Xu, Q. and Cao, L.},
  title   = {{PatentScore}: Multi-dimensional evaluation of {LLM}-generated patent claims},
  journal = {arXiv preprint arXiv:2505.19345},
  year    = {2025}
}

@article{yoo2025sfd,
  author  = {Yoo, Y. and Zhang, X. and Cao, L.},
  title   = {Self-filtered distillation with {LLMs}-generated trust indicators for reliable patent classification},
  journal = {arXiv preprint arXiv:2510.05431},
  year    = {2025}
}

@article{fischer2014,
  author  = {Fischer, T. and Leidinger, J.},
  title   = {Testing patent value indicators on directly observed patent value---An empirical analysis of Ocean Tomo patent auctions},
  journal = {Research Policy},
  volume  = {43},
  number  = {3},
  pages   = {519--529},
  year    = {2014}
}

@article{kalip2022,
  author  = {Kalip, N. G. and Erzurumlu, Y. \"{O}. and G\"{u}n, N. A.},
  title   = {Qualitative and quantitative patent valuation methods: A systematic literature review},
  journal = {World Patent Information},
  volume  = {69},
  pages   = {102111},
  year    = {2022}
}

@article{han2015,
  author  = {Han, E. J. and Sohn, S. Y.},
  title   = {Patent valuation based on text mining and survival analysis},
  journal = {Journal of Technology Transfer},
  volume  = {40},
  number  = {5},
  pages   = {821--839},
  year    = {2015}
}

@article{gambardella2007,
  author  = {Gambardella, A. and Giuri, P. and Luzzi, A.},
  title   = {The market for patents in Europe},
  journal = {Research Policy},
  volume  = {36},
  number  = {8},
  pages   = {1163--1183},
  year    = {2007}
}

@article{gans2008,
  author  = {Gans, J. S. and Hsu, D. H. and Stern, S.},
  title   = {The impact of uncertain intellectual property rights on the market for ideas: Evidence from patent grant delays},
  journal = {Management Science},
  volume  = {54},
  number  = {5},
  pages   = {982--997},
  year    = {2008}
}

@article{akerlof1970,
  author  = {Akerlof, G. A.},
  title   = {The market for ``lemons'': Quality uncertainty and the market mechanism},
  journal = {The Quarterly Journal of Economics},
  volume  = {84},
  number  = {3},
  pages   = {488--500},
  year    = {1970}
}

@article{pakes1989,
  author  = {Pakes, A. and Simpson, M.},
  title   = {Patent renewal data},
  journal = {Brookings Papers on Economic Activity: Microeconomics},
  volume  = {1989},
  pages   = {331--410},
  year    = {1989}
}

@article{schankerman1986,
  author  = {Schankerman, M. and Pakes, A.},
  title   = {Estimates of the value of patent rights in European countries during the post-1950 period},
  journal = {The Economic Journal},
  volume  = {96},
  number  = {384},
  pages   = {1052--1076},
  year    = {1986}
}

@article{abbasiantaeb2025,
  author  = {Abbasiantaeb, Z. and Verberne, S. and Wang, J.},
  title   = {Tracing science-technology-linkages: A machine learning pipeline for extracting and matching patent in-text references to scientific publications},
  journal = {Information Processing \& Management},
  volume  = {62},
  number  = {6},
  pages   = {104264},
  year    = {2025}
}

@article{fu2025,
  author  = {Fu, Z. and Zhu, Q. and Liu, B. and Yan, C.},
  title   = {Patent lifespan prediction and interpreting the key determinants: An application of interpretable machine learning survival analysis approach},
  journal = {Technological Forecasting and Social Change},
  volume  = {215},
  pages   = {124104},
  year    = {2025}
}

@article{jiang2025,
  author  = {Jiang, L. and Goetz, S. M.},
  title   = {Natural language processing in the patent domain: a survey},
  journal = {Artificial Intelligence Review},
  volume  = {58},
  number  = {7},
  pages   = {214},
  year    = {2025}
}

@article{kalip2025,
  author  = {Kalip, N. G. and Bozbura, F. T. and Atteyih, Y.},
  title   = {A comprehensive framework for patent value indicators: Machine learning-driven value prediction in renewable energy patents},
  journal = {World Patent Information},
  volume  = {81},
  pages   = {102364},
  year    = {2025}
}

@article{lee2025,
  author  = {Lee, J. S.},
  title   = {{InstructPatentGPT}: training patent language models to follow instructions with human feedback},
  journal = {Artificial Intelligence and Law},
  volume  = {33},
  number  = {3},
  pages   = {739--782},
  year    = {2025}
}

@article{liu2024a,
  author  = {Liu, J. and Li, P. and Liu, X.},
  title   = {Patent lifetime prediction using {LightGBM} with a customized loss},
  journal = {PeerJ Computer Science},
  volume  = {10},
  pages   = {e2044},
  year    = {2024}
}

@article{schmitt2025,
  author  = {Schmitt, V. J.},
  title   = {Disentangling patent quality: Using a large language model for a systematic literature review},
  journal = {Scientometrics},
  volume  = {130},
  number  = {1},
  pages   = {267--311},
  year    = {2025}
}

@inproceedings{shomee2025,
  author    = {Shomee, H. H. and Wang, Z. and Ravi, S. N. and Medya, S.},
  title     = {A survey on patent analysis: From {NLP} to multimodal {AI}},
  booktitle = {Proceedings of the 63rd Annual Meeting of the Association for Computational Linguistics (Volume 1: Long Papers)},
  pages     = {8545--8561},
  year      = {2025}
}

@article{xiong2025,
  author  = {Xiong, S. and Chen, S. and He, J. and Liu, Y. and Mao, J. and Liu, C.},
  title   = {Scalable multi-label patent classification via iterative large language model-assisted active learning},
  journal = {World Patent Information},
  volume  = {82},
  pages   = {102380},
  year    = {2025}
}

@article{yoshikawa2025,
  author  = {Yoshikawa, N. and Krestel, R.},
  title   = {Do large language models understand patents? {E}nhancing patent classification through {AI}-generated summaries},
  journal = {World Patent Information},
  volume  = {81},
  pages   = {102353},
  year    = {2025}
}

@article{yan2025,
  author  = {Yan, B. and Li, K. and Xu, M. and Dong, Y. and Zhang, Y. and Ren, Z. and Cheng, X.},
  title   = {On protecting the data privacy of Large Language Models ({LLMs}) and {LLM} agents: A literature review},
  journal = {High-Confidence Computing},
  pages   = {100300},
  year    = {2025}
}

@article{jiang2026,
  author  = {Jiang, T. and Xu, Y.},
  title   = {Human-like or machine-like? {H}ow anthropomorphic framing shapes older adults’ attitudes toward health {AI}},
  journal = {Information Processing \& Management},
  volume  = {63},
  number  = {3},
  pages   = {104489},
  year    = {2026}
}

@article{liu2023,
  author  = {Liu, W. and Li, S. and Cao, Y. and Wang, Y.},
  title   = {Multi-task learning based high-value patent and standard-essential patent identification model},
  journal = {Information Processing \& Management},
  volume  = {60},
  number  = {3},
  pages   = {103327},
  year    = {2023}
}

@article{yun2022,
  author  = {Yun, S. and Cho, W. and Kim, C. and Lee, L.},
  title   = {Technological trend mining: Identifying new technology opportunities using patent semantic analysis},
  journal = {Information Processing \& Management},
  volume  = {59},
  number  = {4},
  pages   = {102993},
  year    = {2022}
}

@article{porter2023,
  author  = {Alan L. Porter and Michael Markley and Robert Snead and Nils C. Newman},
  title   = {Twenty years of {US} nanopatenting: Maintenance renewal scoring as an indicator of patent value},
  journal = {World Patent Information},
  volume  = {73},
  pages   = {102178},
  year    = {2023}
 }

@article{jou2018,
  author  = {Jing Ben Jou},
  title   = {{R\&D} investment and patent renewal decisions},
  journal = {The Quarterly Review of Economics and Finance},
  volume  = {69},
  pages   = {144--154},
  year    = {2018}
 }

@article{danish2020,
  author  = {Mir Sayed Shah Danish and Pooja Ranjan and Ritu Sharma},
  title   = {Valuation of patents in emerging economies: A renewal model-based study of Indian patents},
  journal = {Technology Analysis \& Strategic Management},
  volume  = {32},
  number  = {4},
  pages   = {457--473},
  year    = {2020},
  }

\end{document}